\documentclass[12pt,draftclsnofoot,onecolumn,peerreview,journal]{IEEEtran}
\usepackage{graphicx}
\usepackage{verbatim,flafter,caption,amsmath,amssymb}
\usepackage{mathrsfs}

\makeatletter
\def\fps@figure{hbp}
\makeatother
\makeatletter
  \newcommand\figcaption{\def\@captype{figure}\caption}
  \newcommand\tabcaption{\def\@captype{table}\caption}
\makeatother

\begin{document}
\title{Hybrid Decoding of Finite Geometry LDPC Codes}
\author{Guangwen Li, Dashe Li, Yuling Wang, Wenyan Sun }
\maketitle
\bibliographystyle{ieeetran}
\begin{abstract}
For finite geometry low-density parity-check codes, heavy row and
column weights in their parity check matrix make the decoding with
even Min-Sum (MS) variants computationally expensive. To alleviate
it, we present a class of hybrid schemes by concatenating a parallel
bit flipping (BF) variant with an Min-Sum (MS) variant. In most SNR
region of interest, without compromising performance or convergence
rate, simulation results show that the proposed hybrid schemes can
save substantial computational complexity with respect to MS variant
decoding alone. Specifically, the BF variant, with much less
computational complexity, bears most decoding load before resorting
to MS variant. Computational and hardware complexity is also
elaborated to justify the feasibility of the hybrid schemes.
\end{abstract}
\IEEEpeerreviewmaketitle
\section{Introduction}
Low-density parity-check (LDPC) codes, given a sufficiently long
block length, can approach Shannon limit with belief propagation
(BP) decoding \cite{mackay1999gec}\cite{chung2001dld}. Hence, it
remains a research focus among others in the coding field. Lately, a
class of finite geometry (FG) LDPC codes have attracted great
interest, by virtue of the fact that they are encodable in linear
time with feedback shift registers
\cite{kou2001ldp}\cite{tang2005cfg}. However, compared to other
classical LDPC codes, it require much more complexity to decode with
standard BP algorithms for FG-LDPC codes, due to heavy row and
column weights in their parity check matrix.

There exist many low complexity decoding schemes applicable for
FG-LDPC codes. The hard decodings
\cite{gallager1962ldp}\cite{lucas2000ido} have the least complexity
but suffer severe performance loss. To alleviate the degradation, at
the cost of moderate complexity, a class of bit flipping (BF)
variants improve performance after taking into account the soft
information of received sequences. In \cite{Liu2005}, a BF function
was devised wherein both the most and the least reliable bits
involved in one check sum are considered. Further improvement was
reported \cite{shan2005iwb} by weighting each term in the BF
function. A bootstrapping step \cite{Nouh2002}\cite{nouh2004rbs} was
proposed to update those unreliable bits prior to calculating their
BF function values. Based on \cite{kou2001ldp}, the methods
presented in \cite{zhang2004mwb}\cite{jiang2005imw} achieved better
performance, as a result of considering the impact of its received
soft information on the BF function value of a specific bit.
However, one common drawback of above BF variants is that only one
bit is flipped per iteration, which is adverse to fast convergence
requirement. To lower the decoding latency caused by such serial BF
strategy, \cite{wu2007pwb}, \cite{ngatched2009ida} and
\cite{li2008hid} presented three decoding methods in the form of
multi-bit flipping per iteration. In \cite{wu2007pwb}, when the
flipping signal counter for each bit has reached a predesigned
threshold, the pertinent bits flip immediately; in
\cite{ngatched2009ida}, the number of bits chosen to be flipped
approximates the quotient of the number of unsatisfied check sums
and the column weight of parity check matrix. In  \cite{li2008hid},
it was suggested to flip those bits with positive flipping function
values per iteration. Further decoding gain is obtained by adding
into these multi-bit flipping algorithms a delay-handling procedure
\cite{li2009ipw}, which delays flipping those bits whose soft
information presents higher magnitude among others. With respect to
the serial flipping, these parallel or multi-bit flipping methods
show a significant convergence advantage at no cost of performance
loss.

On the other hand,  substantial complexity is saved by estimating
complex $tanh$ function in standard BP with simple $min$ function,
which leads to Min-Sum (MS) or BP-based algorithm
\cite{wiberg1996cad}\cite{fossorier1999rci}. Then MS variants such
as normalized Min-Sum (NMS) and offset Min-Sum (OMS)
\cite{chen2005rcd} proves effective to fill most performance gap
between MS and standard BP, at the cost of minor complexity
increase.

Despite this, the heavy row and column weights of FG-LDPC codes may
annoy the MS variants from perspective of complexity; while the BF
variants present much less complexity but suffering some performance
loss. To expect good performance and low complexity simultaneously,
one natural way is to concatenate some component decoders to fulfill
one decoding. This strategy was attempted in \cite{li2008hid},
wherein standard BP is called only when a multi-bit BF scheme
failed. However, due to the serious performance mismatch between
standard BP and the multi-bit BF scheme proposed in
\cite{li2008hid}, such concatenation results in frequent invocations
of standard BP in most of waterfall SNR region, which subsequently
weakens the efforts of reducing complexity. Different from it, a
gear shift decoding was presented in \cite{ardakani2006gsd}, it
selects appropriate decoder among available ones at each iteration,
according to the optimal trellis route obtained after extrinsic
information transfer (EXIT) chart analysis. Theoretically, the gear
shift decoding reaches the targets of reducing decoding latency
while keeping performance. But several obstacles hinder its
application for finite FG-LDPC codes. For one thing, the delicate
optimal decoding route derived from EXIT chart analysis may deviate
seriously from the real situation, since EXIT chart analysis is
accurate largely for codes of large girth but FG-LDPC codes are
known for the existence of abundant short loops. For another, the
EXIT chart of BF variants remains unknown, but excluding such a
class of decoding schemes may lead to an absence of a competitive
decoder option for gear shift decoding.

In the paper, we adopt a similar framework to that of
\cite{li2008hid} where two two component decoders form an hybrid
scheme.  The former component decoder may be substituted with a
newly proposed BF variant; the latter is an MS variant instead of
standard BP, considering near BP performance is achieved with such
an MS variant. At modest and high SNR regions, both decoding
performance of the latter decoder and low computational complexity
close to the former decoder are achieved, which are verified via
simulations and complexity analysis.

The remainder of the paper is organized as follows. Section
\makeatletter \@Roman{2} \makeatother discusses the motivation of
designing such a class of hybrid schemes. Section \makeatletter
\@Roman{3} \makeatother describes its implementation using BF and MS
variants. Simulation results, convergence rate and complexity
analysis are presented in Section \makeatletter \@Roman{4}
\makeatother. Finally Section \makeatletter \@Roman{5} \makeatother
concludes the work.

\section{Motivation of hybrid decoding}
With the goals of high performance and low
complexity, a satisfying
concatenation of two component decoders meets four conditions. First, the two decoders
present distinct characteristics. Specifically, the former requires much less complexity than the
latter. Secondly, the performance gap between them is within some limit to ensure
performance match. In other words, while no gap wipes off the need of employing hybrid
schemes, excessively large gap, manifested by no well overlapped waterfall regions for both decoders,
results in frequent invocations of the second component decoder. Thirdly, in order not to worsen the whole decoding latency,
it is beneficial that the convergence rates of two decoders are comparable by and large.
Lastly, the hardware complexity of both decoders is shared to the greatest extent to
lower implementation cost.

In \cite{li2008hid}, a multi-bit flipping scheme and standard BP are jointed to
serve the purpose of decoding. However, the multi-bit flipping suffers serious
performance loss when compared to standard BP. Such a concatenation
 violates the mentioned condition two and is less meaningful, since
 standard BP still takes a substantial load in most SNR region. Compared to
the serial ones, the multi-bit BF variants requires
much less decoding iterations \cite{li2009ipw}.
According to condition three, multi-bit BF variant is thus preferable over serial one
when selecting the first component decoder. Moreover, the multi-bit BF variant with
the least complexity and the closest performance to its successor has the highest priority.
On the other hand, for FG-LDPC codes, MS variants with proper correcting factors,
present almost the same performance as standard BP, thus good candidates of
the second component decoder.

\section{Implementation of hybrid decoders}
Assume a binary $(N,K)$ LDPC code with block length $N$ and
dimension $K$. Its parity check matrix is of the form
$\mathbf{\mathbf{H}}_{M\times{N}}$, where $M$ is the number of check
sums. For high rate FG-LDPC codes, the relation $M=N$ indicates
there exist many redundant check sums in $\mathbf{H}$. The BPSK
modulation maps a codeword
$\mathbf{\mathbf{c}}=[c_1,c_2,\ldots,c_N]$ to a symbol sequence
$\mathbf{\mathbf{x}}=[x_1,x_2,\ldots,x_N]$ with $x_i=1-2c_i$, where
$i=1,2,\ldots,N$. After the symbols are transmitted through an
additive white Gaussian noise (AWGN) memoryless channel, we obtain
at the receiver a corrupted sequence
$\mathbf{\mathbf{y}}=[y_1,y_2,\ldots,y_N]$ , where $y_i=x_i+z_i$,
$z_i$ is an independent Gaussian random variable with zero mean and
variance $\sigma^2$.

For convenience, the vectors below are treated as column or row
vectors depending on the context. To differentiate each BF variant,
the initials of the first two authors' surname hyphened by the
letters "WBF" make up a unique name, unless stated otherwise.
\subsection{BF variants}
In LP-WBF \cite{Liu2005}, the BF function of variable node $i$ at
the $l$-th iteration is defined as
\begin{equation}
\label{flipping_function1}
f_i^{(l)}=\sum_{k\in
\mathcal{M}(i)}f_{i,k}^{(l)}, \; i \in [1,N]
\end{equation}
\begin{equation}
\label{flipping_function1.1}
f_{i,k}^{(l)}=\left\{\begin{array}{ll}
|y_i|-\frac{1}{2}(\min_{j \in \mathcal{N}(k)}|y_j|)&\text{if } s_k^{(l)}=0,\\
|y_i|-\frac{1}{2}(\min_{j \in \mathcal{N}(k)}|y_j|)-\max_{j \in
\mathcal{N}(k)}|y_j|&\text{if } s_k^{(l)}=1.\end{array} \right.
\end{equation}
where $\mathcal{M}(i)$ denotes the neighboring check nodes of variable
node $i$, $\mathcal{N}(k)$ is the neighboring variable nodes of check
node $k$, $s_k^{(l)}$ is the $k$-th component of syndrome
$\mathbf{s}$ at the $l$-th iteration.

With the intuition that the more reliable bits involved in a check
sum, the more reliable the check will be, SZ-WBF \cite{shan2005iwb}
defines a BF function by weighting each term of the summation
(\ref{flipping_function1}). That is,
\begin{equation}
\label{flipping_function2}
f_i^{(l)}=\sum_{k\in
\mathcal{M}(i)}w_{i,k}f_{i,k}^{(l)}, \; i \in [1,N]
\end{equation}
\begin{equation}
\label{flipping_function2.1}
w_{i,k}=\max(0,\alpha_1-\|\{j|\,|y_j|\leq \beta_1, j \in
\mathcal{N}(k)\backslash i\}\|)
\end{equation}
where $\mathcal{N}(k)\backslash i$ denotes the neighboring variable
nodes of check node $k$ except variable node $i$, $\alpha_1$ is an
integer constant, $\beta_1$ is a real constant, $\|\cdot\|$ is to
obtain the set cardinality.

For serial BF variants such as SZ-WBF, only one bit with the smallest
$f_i^{(l)}$ is flipped at the $l$-th iteration. Hence, the
maximum number of iterations $I_m$ needs to be predesigned high
enough to allow decoding convergence.

Due to a positive correlation between the number of erroneous bits
and that of unsatisfied check sums, NT-WBF \cite{ngatched2009ida}
suggests flipping $\lambda^{(l)}$ bits of the smallest $f_i^{(l)}$
defined by (\ref{flipping_function1}) at the $l$-th iteration,
$$
\lambda^{(l)}=\lfloor \frac{w_h(\mathbf{s}^{(l)})}{d_v}\rfloor
$$
where $w_h(\cdot)$ denotes the calculation of Hamming weight, $d_v$
is the column weight of matrix $\mathbf{H}$, $\lfloor x \rfloor$ is
the integral part of $x$.

At each iteration, LZ-WBF \cite{li2008hid} flips all the bits with flipping function values greater than zero, among
which, the flipping function is defined as \cite{zhang2004mwb}
\begin{equation}
\label{flipping_function_LZ}
f_i^{(l)}=\sum_{k\in
\mathcal{M}(i)}(2s_k^{(l)}-1)(\min_{j\in N(k)}|y_j|)-\beta_2|y_i|, \; i \in [1,N]
\end{equation}
where $\beta_2$ is a real weighting factor.

WZ-WBF \cite{wu2007pwb} uses the same BF function as in
\cite{jiang2005imw}, namely,
\begin{equation}
\label{flipping_function3}
f_i^{(l)}=\sum_{k\in
\mathcal{M}(i)}(2s_k^{(l)}-1)(\min_{j\in N(k)\backslash i}|y_j|)-\beta_3|y_i|, \; i \in [1,N]
\end{equation}
where $\beta_3$ is a real weighting factor. Then at each iteration, for
each unsatisfied check sum, a flipping signal is assigned to some
involved bit. And only those bits are flipped which have accumulated
flipping signals more than a threshold.

To prevent some reliable bits from flipping hastily, improved parallel weighted bit flipping
(IPWBF) \cite{li2009ipw} added a delay-handling step into the steps
of WZ-WBF.

Compared with IPWBF, the proposed LF-WBF varies by utilizing the
BF function of SZ-WBF, while keeping other steps largely unchanged.
To be self-contained, LF-WBF is described as follows:
\begin{enumerate}
\item
Preprocess:  Assume a threshold $T$ be the value of the
$\lfloor\beta_4N\rfloor$-th smallest element among array $|y_i|,
i\in [1,N]$, where $\beta_4$ is a real constant within $[0,1]$, then those
bits with $|y_i|$ greater than $T$ are marked reliable, otherwise
unreliable.
\item
Initialize: $l\leftarrow0$; calculate initial values of $f_i^{(0)}$,
$i\in [1,N]$ according to (\ref{flipping_function1.1}),
(\ref{flipping_function2}). For the bits $\in\{i||y_i|>T,
i\in[1,N]\}$, the delay-handling counters $a_i\leftarrow0$; note
hard-decision of $\mathbf{y}$ as $\hat{\mathbf{c}}^{(0)}$.
\item
Syndrome and reset: Calculate
$\mathbf{s}^{(l)}=\mathbf{H}\hat{\mathbf{c}}^{(l)}$. If
$\mathbf{s}^{(l)}=\mathbf{0}$, stop to return
$\hat{\mathbf{c}}^{(l)}$ as the decoding result. If not,
$b_i\leftarrow0, i\in [1,N]$, $b_i$ is a flipping counter which sums
the flipping signals for bit $i$.
\item
Collect flipping signals: Update $f_i^{(l)}$, $i\in [1,N]$ based on
(\ref{flipping_function1.1}), (\ref{flipping_function2}). For each
$k\in\{k|s_k^{(l)}\neq0, k\in [1,M]\}$, identify the index
$i^*=\arg\min_{i\in \mathcal{N}(k)}f_i^{(l)}$, then
$b_{i^*}\leftarrow b_{i^*}+1$, that is, a flipping signal is
collected for bit $i^*$.
\item
Decide flipping bits: It is divided into two substeps.
\begin{enumerate}
\item
For the bits $\in\{i|b_i\geq\alpha_2, i\in [1,N]\}$, where
$\alpha_2$, as a positive integer,  represents the flipping
threshold, flip them if only the resulting syndrome
$\mathbf{s}^{(l+1)}=\mathbf{0}$. Otherwise turn to the next substep.
\item
Delay-handling: For the unreliable bits $\in\{i|b_i\geq\alpha_2,
|y_i|\leq T, i\in [1,N]\}$, put them in a to-be-flipped list; for
the reliable bits $\in\{i|b_i\geq\alpha_2, |y_i|>T, i\in [1,N]\}$,
update by $a_i\leftarrow a_i+1$. Subsequently, put the bits
$\in\{i|a_i\geq\alpha_3, i\in [1,N]\}$ in the to-be-flipped list,
where $\alpha_3$ is a small positive integer defining a
delay-handling threshold. Obviously, it is meaningful only for
$\alpha_3\geq 2$. Relax $\alpha_3\leftarrow \alpha_3-1$ if only the
to-be-flipped list is empty, then flip the bits
$\in\{i|b_i=\alpha_3, i\in [1,N]\}$. Declare failure if no bit is
qualified yet.

\end{enumerate}
 Since delay-handling step may potentially increase the average number of decoding
 iterations,  substep one  reduces its impact effectively.
 \item
Flip and reset: Flip these bits in the to-be-flipped list. Reset all
the bits $\in\{i|a_i\geq\alpha_3, i\in [1,N]\}$ by $a_i\leftarrow0$.
Noticeably, before the next resetting occurs, the duration of $a_i$
may last several iterations while that of $b_i$ is always one
iteration.
\item
$l\leftarrow l+1$. If $l<I_m$, goto step $3$ to continue one more
iteration; otherwise, declare failure.
\end{enumerate}

\subsection{MS variants}
At the check nodes end, compared with standard BP implemented in
Log-likelihood ratio (LLR) domain, NMS and OMS \cite{chen2005rcd},
approximate (\ref{LLR_check}) with (\ref{min_sum_expression1}) and
(\ref{min_sum_expression2}), respectively, thus saving most
complexity,
\begin{equation}
\label{LLR_check}
L^{(l)}_{j,i}=2\tanh^{-1}\Big(\prod_{k\in \mathcal {N}(j)\backslash i}\tanh \frac{Z^{(l-1)}_{j,k}}{2}\,\Big)\\
\end{equation}
\begin{equation}
\label{min_sum_expression1}
\vspace{0.2cm}
L^{(l)}_{j,i}=\frac{1}{\beta_5}\prod_{k\in \mathcal {N}(j)\backslash i}\text{sgn}(Z^{(l-1)}_{j,k})\cdot\min_{k\in \mathcal {N}(j)\backslash i}{|Z^{(l-1)}_{j,k}|}
\end{equation}
\begin{equation}
\label{min_sum_expression2}
\vspace{0.2cm}
L^{(l)}_{j,i}=\prod_{k\in \mathcal {N}(j)\backslash
i}\text{sgn}(Z^{(l-1)}_{j,k})\cdot\max\,(\min_{k\in \mathcal {N}(j)\backslash i}{|Z^{(l-1)}_{j,k}|-\beta_6,0}\,)
\end{equation}
where $L^{(l)}_{j,i}$ denotes the message sent from check node $j$
to variable node $i$ at the $l$-th iteration; $Z^{(l-1)}_{j,k}$
denotes the message sent from variable node $k$ to check node $j$ at
the $(l-1)$-th iteration; $\beta_5$ or $\beta_6$, being a real constant, functions
as a scaling or offset factor, respectively.

To further reduce complexity, at the variable node end, the
calculating of (\ref{LLR_bit_update1}) is approximated with
(\ref{LLR_bit_udpate2}) in the normalized APP-based (NAB) algorithm
\cite{chen2005rcd}.
\begin{equation}
\label{LLR_bit_update1}
Z^{(l)}_{j,i}=F_i+\sum_{k\in \mathcal {M}(i)\backslash j}L^{(l)}_{k,i}\\
\end{equation}
\begin{equation}
\label{LLR_bit_udpate2}
Z^{(l)}_{j,i}=F_i+\sum_{k\in \mathcal {M}(i)}L^{(l)}_{k,i},\,\forall j \in \mathcal {M}(i)\\
\end{equation}
Where $F_i$ is the initial LLR of bit $i$. For the difference-set
cyclic (DSC) codes, it was reported NAB yields almost as good
performance as NMS \cite{chen2005rcd}. As shown in the simulation
later, similar observation also holds for FG-LDPC codes.
\subsection{Block graph of a hybrid decoding scheme}
\begin{figure}
\centering
\includegraphics[width=0.75\linewidth]{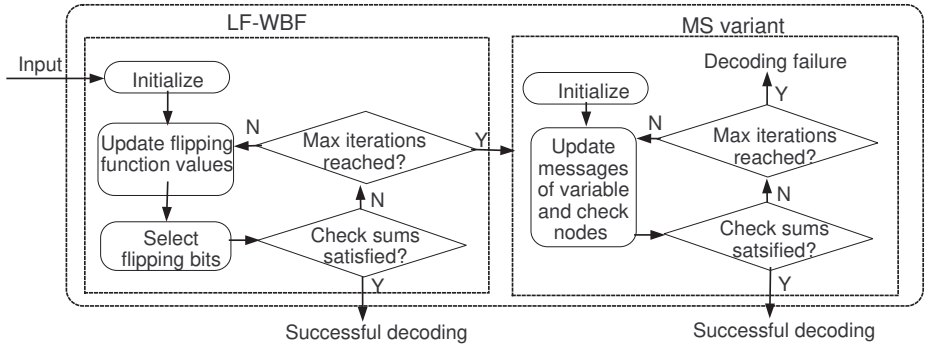}
\parbox{0.9\linewidth}{\caption{\label{fig:block_gram} Block graph of hybrid decoding scheme}}
\end{figure}
There are many BF variants and MS variants, thus the combinations of BF variant plus MS variant
is abundant. For instance of 'LF-WBF+NMS', as shown in \ref{fig:block_gram}, two component decoders are independent comparatively. The latter
takes over decoding  so long as the former  failed.
\subsection{Optimize parameters by differential evolution}
It is hard to optimize the group of parameters involved in LF-WBF
theoretically. Hence, differential evolution (DE), known as a
heuristic search method, is exploited to approximate the optimality.
Similar to the genetic algorithm, DE is a simple and reliable
optimization tool \cite{storn1997des}. In DE, via various operations
including  mutation, combination and selection, a population of
solution vectors  are updated generation by generation, with those
new vectors with small objective values survived, until the
population converges to the global optimum.

To aid LF-WBF to optimize its parameter vector
$(\alpha_1,\alpha_2,\alpha_3,\beta_1,\beta_4)$, the objective
function of DE is designated to find the minimum bit error rate (BER)
given a block of received sequences. In order to save computation, each
parameter of LF-WBF is roughly assigned an evaluation interval
beforehand. For instance, $\alpha_1, \alpha_2$ are integers in
$[1,d_v/2]$, $\alpha_3$ is a small positive in $[1, 4]$, $\beta_1,
\beta_4$ are real numbers in $[0, 1]$.

For $(273,191)$ and $(1023,781)$ FG-LDPC codes \cite{kou2001ldp}, DE
results are given in \ref{parameter_setting} with
varied channel variance $\sigma^2$.

\begin{table}
\centering
\parbox{0.8\linewidth}{\caption{\label{parameter_setting} :
Parameters optimization of LF-WBF for (273,191)\,(left) and (1023,781)\,(right) FG-LDPC codes using differential evolution}}
\begin{tabular}{|c|c|c|c|c|c|}
\hline
$\sigma$& $\alpha_1$ & $\alpha_2$&$\alpha_3$&$\beta_1$&$\beta_4$\\
\hline
0.58&10&4&4&0.31&0.064\\
0.575&9&4&3&0.57&0.11\\
0.57&6&4&4&0.50&0.054\\
0.565&5&3&4&0.47&0.07\\
\hline
\end{tabular}
\hspace{10pt}
\begin{tabular}{|c|c|c|c|c|c|}
\hline
$\sigma$& $\alpha_1$ & $\alpha_2$&$\alpha_3$&$\beta_1$&$\beta_4$\\
\hline
0.565&5&9&2&0.38&0.036\\
0.56&12&8&3&0.51&0.071\\
0.555&10&8&3&0.41&0.075\\
0.55&6&6&3&0.32&0.025\\
\hline
\end{tabular}
\end{table}

\section{Simulation Results and Discussion}
\subsection{Parameters selection}
It is verified that decoding performance of LF-WBF is largely
insensitive to the minor change of its parameters, thus in all SNR
region, we assume the settings as shown on the first row of
\ref{parameter_setting2}, after referring to \ref{parameter_setting}.
The additional advantage of such simplification is that the overall hybrid
decoding requires no more a priori information about the channel, namely, holding as well the property
of uniformly most powerful (UMP)\cite{fossorier1999rci} for MS
variants.

For LZ-WBF and WZ-WBF, the data presented in
\ref{parameter_setting2} come from the existing literature, as
mentioned in the last column of \ref{parameter_setting2}.

After applying DE for MS variants, we select the settings as the
last three rows of \ref{parameter_setting2} for NAB, NMS and OMS.
Noticeably, the optimization results of the scaling factor for NAB
and NMS are different.

\begin{table}
\centering
\parbox{0.6\linewidth}{\caption{\label{parameter_setting2} :
Parameters settings of various decoding schemes for (273,191) and
(1023,781) FG-LDPC codes }}
\begin{tabular}{|c|c|c|}
\hline
Scheme& Parameter(s)=those for (273,191); those for (1023,781)&Source\\
\hline
LF-WBF&$(\alpha_1,\alpha_2,\alpha_3,\beta_1,\beta_4)=(6,4,2,0.45,0.07); (8,7,2,0.4,0.04)$&DE\\
LZ-WBF&$\beta_2$=1.5; 2.1&\cite{li2008hid}\\
WZ-WBF&$(\alpha_2,\beta_3)=(4, 1.3); (10,1.8)$&\cite{xiaofu7tuw}\\
SZ-WBF&$(\alpha_1,\beta_1)$=N/A; (9,0.5)&\cite{shan2005iwb}\\
NAB&$\beta_5$=5.7; 7.1&DE\\
NMS&$\beta_5$=2.9; 3.7&DE\\
OMS&$\beta_6$=0.22; 0.20&DE\\
\hline
\end{tabular}
\end{table}
\subsection{Decoding performance}
\begin{figure}
\centering
\includegraphics[width=0.7\linewidth]{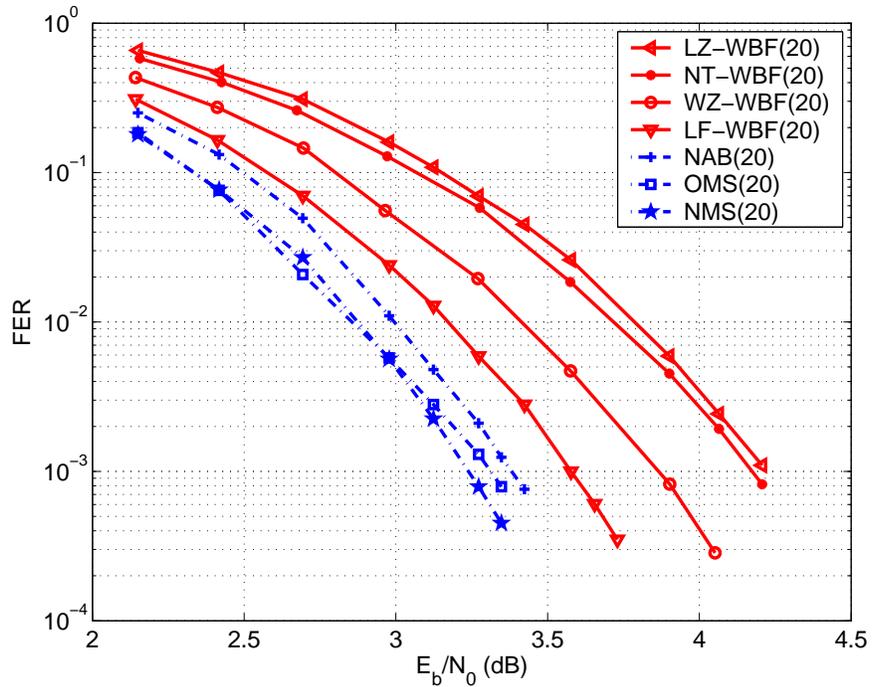}
\parbox{0.9\linewidth}{\caption{\label{fig:273_191_errors} FER curves for (273,191) FG-LDPC  code
under various BF or MS variants}}
\end{figure}

The frame error rate (FER) curves of some BF variants and MS variants  are
plotted in \ref{fig:273_191_errors} for (273,191) code.

In the legend, the number in the brackets stands for the maximum
number of iteration $I_m$. It is found that BF variants are in
general inferior to MS variants from perspective of
performance. Specifically, at the point FER=$10^{-3}$, LF-WBF
leads WZ-WBF, NT-WBF and LZ-WBF about 0.25, 0.58 and 0.6\,dB, respectively.
But it  lags behind NAB, OMS and NMS about 0.2, 0.26, 0.32\,dB, respectively.
Further comparison between LF-WBF and IPWBF \cite{li2009ipw} shows that they present the similar decoding
performance, thus exchangeable each other.
Therefore, LF-WBF owns the most matched SNR region as that of MS variants among the available BF variants.
Considering LDPC codes commonly have large enough minimum distance, the cases seldom occur
where BF variant results in an undetectable error but MS variant decodes correctly.
On the other hand, there exist a few cases where BF variant works but MS variant fails.
Thus in the form of a BF variant plus an MS variant, the hybrid decoding will keep
at least the same performance as the MS variant alone. However, for each combination,
the matching degree between two component decoders impacts heavily the overall decoding complexity.

\begin{figure} \centering
\includegraphics[width=0.7\linewidth]{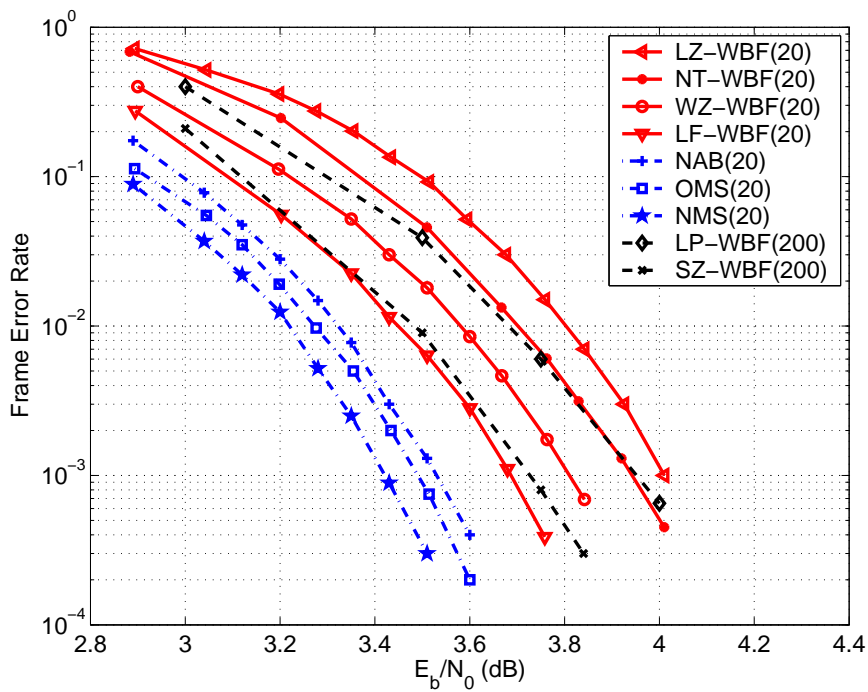}
\parbox{0.9\linewidth}{\caption{\label{fig:1023_781_errors} FER performance for (1023,781) FG-LDPC  code
under BF or MS variants}}
\end{figure}

For (1023,781) code, the FER curves are plotted in \ref{fig:1023_781_errors}.
It is observed that when the block length increases from 273 to 1023,
the curves relation within  BF variants and MS variants still holds, except that
the closeness among these curves  slightly shifts.  For instance, at the point
FER=$10^{-3}$, there exists about $0.3$\,dB between LF-WBF and NMS, while
LF-WBF exceeds LZ-WBF more than 0.3\,dB. Also included are
the curves of two serial approaches: LP-WBF and SZ-WBF.
 The performance of LP-WBF and
SZ-WBF with $I_m=200$ approximates NT-WBF and LF-WBF with $I_m=20$, respectively. Meanwhile, the full loop
detection \cite{shan2005iwb}, which proves
effective in avoiding decoding trappings for serial BF variants, is utilized for
both LP-WBF and SZ-WBF.

\subsection{Convergence rate}
Since some applications require $I_m$ to be small, it is thus meaningful
to investigate the convergence rate of various decoding schemes. At a typical point
SNR=3.42\,dB (or $\sigma$=0.57) of (273,191) code, \ref{convergence_rate} gives performance comparison among
each schemes under varied $I_m$.
\begin{table}
\centering
\parbox{0.7\linewidth}{\caption{\label{convergence_rate} :
Decoding performance of various schemes under varied $I_m$ for (273,191) FG-LDPC code at SNR=3.42\,dB}}
\begin{tabular}{|c|c|c|c|c|c|}
\hline
Scheme& $I_m=3$&$I_m=10$&$I_m=20$&$I_m=50$&$I_m=200$\\
\hline
LZ-WBF&9.2e-2&3.6e-2&3.6e-2&3.6e-2&3.6e-2\\
NT-WBF&5.9e-1&4.4e-2&3.8e-2&3.8e-2&3.8e-2\\
WZ-WBF&2.5e-2&9.6e-3&9.8e-3&9.8e-3&9.8e-3\\
LF-WBF&4.5e-2&4.2e-3&2.8e-3&2.4e-3&2.3e-3\\
NAB&1.1e-1&2.1e-3&7.6e-4&4.4e-4&4.4e-4\\
OMS&1.6e-2&9.6e-4&6.6e-4&5.0e-4&4.6e-4\\
NMS&1.1e-2&5.2e-4&3.8e-4&3.6e-4&3.4e-4\\
BP&9.4e-3&3.9e-3&2.9e-3&1.6e-3&7.2e-4\\
\hline
\end{tabular}
\end{table}

It is seen that although $I_m=3$ is too rigorous for all decoding schemes, each BF variant
reaches its individual decoding capability at the specified point within $I_m=20$. That is,
more iterations after the 20-th iteration achieves no further decoding improvement; while MS
variants require $I_m$ to be at least 50 to fully decode the received sequences.
Also included in \ref{convergence_rate} is the data of BP. Interestingly, at $I_m=3$, BP
yields the best decoding performance among others. But its convergence rate is not satisfying.
It is shown that $I_m=50$ is not even
sufficient for BP decoding, because the performance improves from FER=1.6e-3 to
7.2e-4 when $I_m$ increases to 200. For this reason, given a small $I_m=20$, LF-WBF even excels
BP a little, as shown in \ref{convergence_rate}. Further simulation shows that
 LF-WBF with $I_m=20$ performs better than BP in the region where SNR is greater than 3.42\,dB.
 Another noticeable point shown in \ref{convergence_rate} is that
 the performance of BP is generally inferior to MS variants, despite its high complexity. Therefore, BP is less attractive
 to be selected as the second component decoder of a hybrid scheme.
 Taking into account the fact that serial
 BF variants require much more $I_m$ than above multi-bit BF variants \cite{ngatched2009ida}\cite{li2009ipw},
 and LF-WBF performs the best among existing multi-bit BF variants, LF-WBF plus some MS variant
 intuitively presents a competitive form of hybrid decoding scheme. The similar points
 are supported as well after generalized to other longer FG-LDPC codes.

\begin{figure}
\centering
\includegraphics[width=0.7\linewidth]{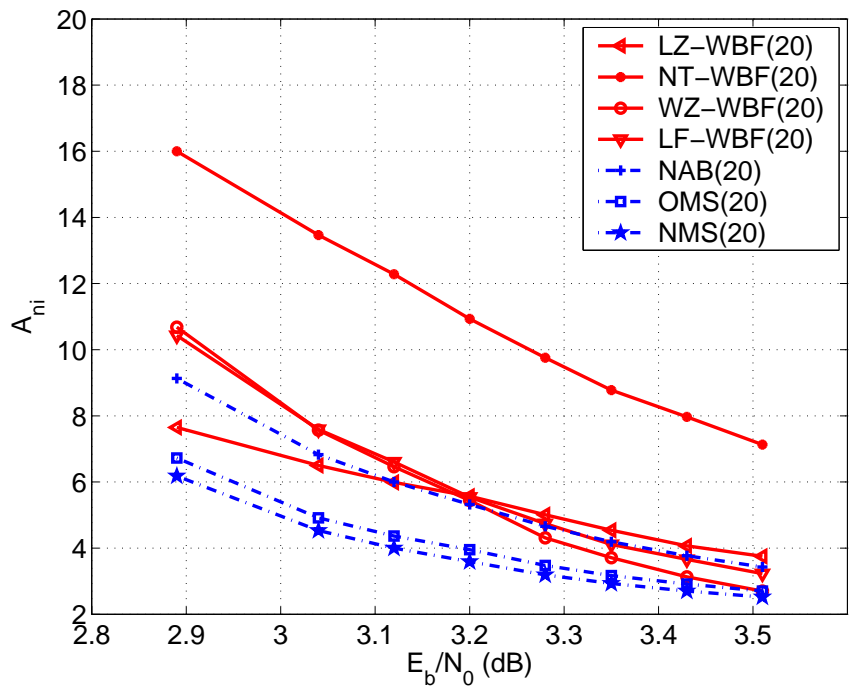}
\parbox{0.9\linewidth}{\caption{\label{fig:iterations} Average number of iterations $A_{ni}$ of various decodings
schemes for (1023,781) FG-LDPC code}}
\end{figure}

Let $A_{ni}$ denote average number of iterations for each decoding scheme, as seen in
\ref{fig:iterations} for (1023,781) code, $A_{ni}$ of NT-WBF sticks out prominently while
that of LZ-WBF varies slowly with $E_b/N_0$, both are due to the algorithms themselves. In most
SNR region of interest, all BF variants except NT-WBF present comparable $A_{ni}$ as MS variants, thus
meeting well  the condition three discussed in Section \makeatletter \@Roman{2} \makeatother.

\subsection{Computational complexity analysis}
Practically, any BF variant followed by an MS variant will yield the same decoding
performance as the latter alone. For instance, LF-WBF plus MS variant performs almost equally as
LZ-WBF plus MS variant, regardless of the fact that LF-WBF is far superior to LZ-WBF.
However, computational complexity differs enormously with respect to each hybrid scheme.
Generally, it is hard to accurately describe the required complexity for each decoding scheme,
so data obtained in the simulations is presented to support our viewpoints if necessary.

Let  $d_v, d_c$  individually denote the column and row weights of
parity check matrix $\mathbf{H}$, then for each BF variant, its complexity roughly consists of three
parts: preprocessing, updating BF function  and selecting flipping bits. To the best of our knowledge,
the complexity of preprocessing and initialization is largely
omitted in existing literature. However, the following analysis and simulation will
show it contributes substantially to the complexity at very high SNR region.
Ignoring simple binary operations and a small number of
real multiplications involved sometimes, it suffice to address the
dominant real additions for each BF variant, assuming one real
comparison is treated as one real addition.

At the stage of preprocessing, for LP-WBF and  NT-WBF, about $N(2d_c-3)$ comparisons are
needed in computing $min$ and $max$ terms
of (\ref{flipping_function1.1}). Similarly, for LZ-WBF and WZ-WBF, about $N(d_c-1)$ comparisons is required individually
in computing  the $min$ term of (\ref{flipping_function_LZ}) and (\ref{flipping_function3}).
Besides that for LP-WBF, both SZ-WBF and LF-WBF require extra $N$ comparisons to obtain $w_{i,k}$ term
of (\ref{flipping_function2.1}). With respect to SZ-WBF, LF-WBF requires
 about $N\log_2\lfloor\beta_4N\rfloor+N$ more comparisons  to determine the bit with the
$\lfloor\beta_4N\rfloor$-th smallest magnitude and to mark the delay-flipping bits.

Prior to updating the BF function of each bit, it is initialized with $d_v-1$ additions for each BF
variant. For multi-bit BF variants, there are two ways of updating the BF function of pertinent bits since the second iteration.
One is to invoke $d_vd_c$ additions per flipped bit to update its BF function;
another is to update the BF function of each bit after
comparing its  column of $\mathbf{H}$ with the syndromes
before and after the latest iteration. The latter is more economical,
considering two flipping bits in the same check sum result in two extra additions for the former,
but avoidable for the latter. For serial BF variants, totally $d_vd_c$ terms
are used to update the BF function of  those affected bits per iteration.

To decide which bits to flip, each BF variant has its own approach.
For LP-WBF and SZ-WBF, it just requires $N-1$ comparisons to find the bit with the smallest
BF function value;  for WZ-WBF and LF-WBF, $d_c-1$ comparisons are required per
unsatisfied check to collect flipping signals for each bit; for NT-WBF, its complexity at this
stage is equal to selecting the smallest say 5 elements in an unordered array. Noticeably, no computation
is required for LZ-WBF, since it simply flips those bits with positive BF function values.

To sum up, \ref{complexity_formulae} gives the complexity composition for
 each BF variant.
\begin{table}
\centering
\parbox{0.7\linewidth}{\caption{\label{complexity_formulae} : Approximated real additions per sequence for
various decoding schemes of FG-LDPC codes}}
\begin{tabular}{|c|c|c|c|} \hline
Schemes&Preprocess&Update BF function (include initialization)&Select bit(s) to flip\\
\hline
LZ-WBF&$N(d_c-1)$&$N(d_v-1)+(A_{ni}-1)NA_{nc}$&0\\
NT-WBF&$N(2d_c-3)$&$N(d_v-1)+(A_{ni}-1)NA_{nc}$&$A_{ni}N\log_2A_{nb}$\\
WZ-WBF&$N(d_c-1)$&$N(d_v-1)+(A_{ni}-1)NA_{nc}$&$A_{ni}A_{ns}(d_c-1)$\\
LF-WBF&$N(2d_c-1+\log_2\lfloor\beta_4N\rfloor)$&$N(d_v-1)+(A_{ni}-1)NA_{nc}$&$A_{ni}A_{ns}(d_c-1)$\\
SZ-WBF&$N(2d_c-2)$&$N(d_v-1)+(A_{ni}-1)d_vd_c$&$A_{ni}(N-1)$\\
LP-WBF&$N(2d_c-3)$&$N(d_v-1)+(A_{ni}-1)d_vd_c$&$A_{ni}(N-1)$\\
\hline
NAB&\multicolumn{3}{|c|}{$A_{ni}(2Nd_v+M(\lceil\log_2d_c\rceil-2))$}\\
OMS, NMS&\multicolumn{3}{|c|}{$A_{ni}(N(4d_v-3)+M(\lceil\log_2d_c\rceil-2))$}\\
\hline
\end{tabular}
\end{table}
In  the table, $A_{nb}$ denotes the average number of selected bits per
iteration for NT-WBF, $A_{nc}$ is the average number of updated
BF function terms per bit per iteration, $A_{ns}$ is the average
number of unsatisfied checks per iteration. Also included are the
complexity expressions of NAB, OMS and NMS as reported in
\cite{Liu2005}, wherein $\lceil\cdot\rceil$ is the ceiling function.

For (1023,781) code, $N=1023, d_v=d_c=32$ \cite{kou2001ldp}. Assume $I_m=20$ for
multi-bit BF variants, $I_m=200$ for MS variants and serial BF variants to ensure full decoding convergence,
at a typical point of SNR=3.28\,dB (or $\sigma=0.555$),
\ref{typical_snr} presents the figures observed in simulation, among which
the last column is the number of real additions according to the expressions
of \ref{complexity_formulae}.
 Noticeably, the last two rows of \ref{typical_snr} gives complexity
of two instances of hybrid decoding schemes as well.
\begin{table}
\centering
\parbox{0.7\linewidth}{\caption{\label{typical_snr} : Complexity comparison per sequence for
various decoding schemes of (1023,781) FG-LDPC code at
SNR=3.28\,dB}}
\begin{tabular}{|c|c|c|c|c|c|} \hline
Scheme& $A_{ni}$ & $A_{ns}$&$A_{nb}$&$A_{nc}$&number of real additions(e+5)\\
\hline
LZ-WBF&4.70&N/A&N/A&8.11&$0.94$\\
NT-WBF&9.61&N/A&9.73&7.72&$1.94$\\
WZ-WBF&4.48&348.01&N/A&10.41&$1.49$\\
LF-WBF&4.74&373.63&N/A&10.10&$1.95$\\
\hline
SZ-WBF&49.08&\multicolumn{3}{|c|}{}&$1.95$\\
LP-WBF&68.66&\multicolumn{3}{|c|}{}&$2.34$\\
NAB&5.53&\multicolumn{3}{|c|}{N/A}&$3.79$\\
OMS&4.47&\multicolumn{3}{|c|}{}&$5.78$\\
NMS&3.77&\multicolumn{3}{|c|}{}&$4.93$\\
\hline
LZ-WBF+NMS&\multicolumn{4}{|c|}{(Data for LZ-WBF) + (NMS with $A_{ni}$=1.88)}&$3.40$\\
LF-WBF+NMS&\multicolumn{4}{|c|}{(Data for LF-WBF) + (NMS with $A_{ni}$=0.88)}&$3.10$\\
\hline
\end{tabular}
\end{table}

After studying \ref{typical_snr}, we find that
the class of BF variants demonstrates a substantial advantage over MS variants
in terms of complexity.  Among the BF variants, LZ-WBF presents the least complexity due to
its fast convergence, low-complexity preprocessing and no complexity demand of selecting bits;
despite its simplicity, at low and modest SNR regions, the combination of LZ-WBF and NMS requires more complexity than
that of  LF-WBF and NMS, as a result that the former combination demands one more
iteration of NMS on average, as shown in \ref{typical_snr}. Under the condition of  offering equivalent performance, the last three rows of
\ref{typical_snr} illustrates that both hybrid decoding schemes can save much complexity, with
respect to its second component decoder alone.

\begin{figure}
\centering
\includegraphics[width=0.7\linewidth]{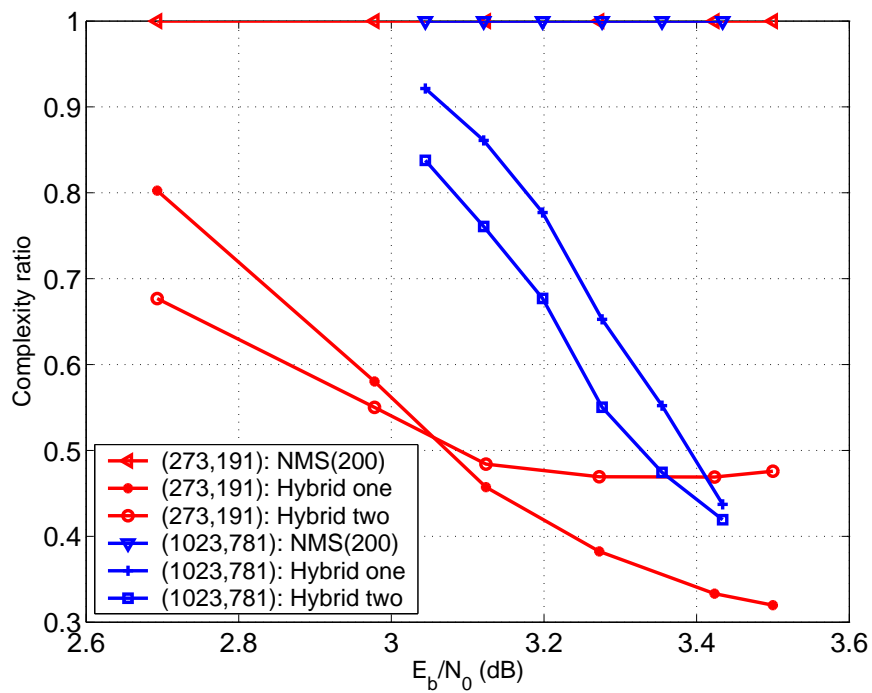}
\parbox{0.8\linewidth}{\caption{\label{fig:performance_ratio}
Complexity ratio curves of Hybrid one: LZ-WBF(20)+NMS(200) and Hybrid two: LF-WBF(20)+NMS(200) for (273,191) and (1023,781) FG-LDPC codes}}
\end{figure}

To better illustrate complexity comparison in the whole SNR region, \ref{fig:performance_ratio}
present complexity ratio curves for (273,191) and (1023,781) codes. Assuming  the complexity of
NMS is a benchmark, then complexity ratio is defined as the ratio of the complexity of a specified
hybrid scheme and that of NMS. For NMS, since another $A_{ni}Nd_v$ divisions is actually required \cite{Liu2005}, we roughly treat as total complexity the sum
of this expression and the related formula in \ref{complexity_formulae}. At very low SNR region,
any of the hybrid schemes shows no much advantage, due to the fact that  most decodings are up to NMS. However, with increased SNR, both hybrid schemes
yield more and more complexity reduction, resulting from more involvements of LZ-WBF or LF-WBF in decoding.
For short (273,191) code, the combination of 'LZ-WBF+NMS' exceeds that of 'LF-WBF+NMS' at the point
SNR=3.05\,dB. While the occurrence  extends to SNR=3.45\,dB for (1023,781) code. Hence, it
suggests that the intersection of these two schemes will move to a higher SNR with
longer block length.

Let $C_{NMS}, C_{LZN}, C_{LFN}$ denote the complexity of above three decoding schemes.
For sufficiently long FG-LDPC codes, to seek
the asymptotic performance ratios in very high SNR region,
the following approximations are derived based on \ref{complexity_formulae},
\begin{equation}
\begin{cases}
\frac{C_{LZN}}{C_{NMS}}=\frac{d_c+d_v-2+(A_{ni}-1)A_{nc}}{A_{ni}(5d_v+\lceil\log_2d_c\rceil-5)}\approx{\frac{2}{5A_{ni}}},\nonumber\\
\frac{C_{LZN}}{C_{NMS}}=\frac{2d_c+d_v-2+\log_2\lfloor\beta_4N\rfloor)+(A_{ni}-1)A_{nc}+A_{ni}(d_c-1)A_{ns}/N}{A_{ni}(5d_v+\lceil\log_2d_c\rceil-5)}\approx{\frac{9+A_{ni}}{15A_{ni}}}.\nonumber

\end{cases}
\end{equation}

wherein the following simulation results are exploited:
$d_v=d_c$, both are large numbers compared with other terms; $A_{ni}$ of various schemes
ranges in [1, 2] and tends to be near each other; $A_{nc}$ of LZ-WBF or LF-WBF is small
compared to $d_v$; $A_{ns}/N$ is about one third.
Similar approach can be used to derive complexity ratios of other hybrid combinations.
\subsection{Hardware complexity}
Seemingly, the proposed hybrid schemes add much more hardware complexity with respect to its second
component decoder alone. However, most hardware complexity can be shared instead between two
component decoders. For instance of 'LF-WBF+NMS', assuming NMS hardware is available, then
$min,max$ operations at the preprocessing phase of LF-WBF, and collecting flipping signals at
the  selecting flipping bits phase of LF-WBF, can be accomplished via the check node logics
of NMS, while the initialization step via the bit node logics of NMS.
Thus compared with NMS, 'LF-WBF+NMS' only includes a few more  integer counters and interconnection logics.
Therefore, the extra hardware complexity of hybrid decoding schemes is largely ignorable.

\section{Conclusions}
For finite FG-LDPC codes, the concatenation of BF variant and MS
variant proves its effectiveness in decoding at a wide rang of SNR
region, by means of achieving performance of the MS variant with
substantial reduced computational complexity.
 While LZ-WBF plus MS variant has its advantage at high SNR region
 of interest; the proposed LF-WBF plus MS variant demonstrates better complexity saving at
 the rest of SNR region, due to the well overlapped waterfall regions between
two component decoders. Evidently, if we can gear among these hybrid
schemes based on varied SNRs, the decoding will be more powerful and
robust.

For BP decoding, it is known that flooding schedule is not optimal.
Sharon et al. \cite{levin2007lsl}\cite{kfir2003pvs}\cite{sharon2007esm} proved
that serial message passing schedule, implemented by fully utilizing
available updated messages, can halve the average number of
iterations of flooding schedule without performance penalty. But it
risks resulting in higher decoding latency. Contrary to it, our
hybrid scheme yields a good tradeoff among performance, complexity
and latency.

\bibliography{IEEEabrv,output1}

\begin{thebibliography}{10}
\providecommand{\url}[1]{#1}
\csname url@rmstyle\endcsname
\providecommand{\newblock}{\relax}
\providecommand{\bibinfo}[2]{#2}
\providecommand\BIBentrySTDinterwordspacing{\spaceskip=0pt\relax}
\providecommand\BIBentryALTinterwordstretchfactor{4}
\providecommand\BIBentryALTinterwordspacing{\spaceskip=\fontdimen2\font plus
\BIBentryALTinterwordstretchfactor\fontdimen3\font minus
  \fontdimen4\font\relax}
\providecommand\BIBforeignlanguage[2]{{%
\expandafter\ifx\csname l@#1\endcsname\relax
\typeout{** WARNING: IEEEtran.bst: No hyphenation pattern has been}%
\typeout{** loaded for the language `#1'. Using the pattern for}%
\typeout{** the default language instead.}%
\else
\language=\csname l@#1\endcsname
\fi
#2}}

\bibitem{mackay1999gec}
D.~MacKay, ``{Good error-correcting codes based on very sparse matrices},''
  \emph{IEEE Transactions on Information Theory}, vol.~45, no.~2, pp. 399--431,
  1999.

\bibitem{chung2001dld}
S.~Chung, G.~Forney~Jr, T.~Richardson, and R.~Urbanke, ``{On the design of
  low-density parity-check codes within 0.0045 dB of the Shannon limit},''
  \emph{IEEE Communications Letters}, vol.~5, no.~2, pp. 58--60, 2001.

\bibitem{kou2001ldp}
Y.~Kou, S.~Lin, and M.~Fossorier, ``{Low-density parity-check codes based on
  finite geometries: arediscovery and new results},'' \emph{IEEE Transactions
  on Information Theory}, vol.~47, no.~7, pp. 2711--2736, 2001.

\bibitem{tang2005cfg}
H.~Tang, J.~Xu, S.~Lin, and K.~Abdel-Ghaffar, ``{Codes on finite geometries},''
  \emph{IEEE Transactions on Information Theory}, vol.~51, no.~2, pp. 572--596,
  2005.

\bibitem{gallager1962ldp}
R.~Gallager, ``{Low-density parity-check codes},'' \emph{IEEE Transactions on
  Information Theory}, vol.~8, no.~1, pp. 21--28, 1962.

\bibitem{lucas2000ido}
R.~Lucas, M.~Fossorier, K.~Yu, and S.~Lin, ``{Iterative decoding of one-step
  majority logic deductible codesbased on belief propagation},'' \emph{IEEE
  Transactions on Communications}, vol.~48, no.~6, pp. 931--937, 2000.

\bibitem{Liu2005}
Z.~Liu and D.~Pados, ``A decoding algorithm for finite-geometry {LDPC} codes,''
  \emph{{IEEE} Transactions on Communications}, vol.~53, no.~3, pp. 415--421,
  Mar. 2005.

\bibitem{shan2005iwb}
M.~Shan, C.~Zhao, and M.~Jiang, ``{Improved weighted bit-flipping algorithm for
  decoding LDPC Codes},'' \emph{IEE Proceedings-Communications}, vol. 152,
  no.~6, pp. 919--922, 2005.

\bibitem{Nouh2002}
A.~Nouh and A.~Banihashemi, ``Bootstrap decoding of low-density parity-check
  codes,'' \emph{{IEEE} Communications Letters}, vol.~6, no.~9, pp. 391--393,
  2002.

\bibitem{nouh2004rbs}
------, ``{Reliability-based schedule for bit-flipping decoding of low-density
  Parity-check codes},'' \emph{IEEE Transactions on Communications}, vol.~52,
  no.~12, pp. 2038--2040, 2004.

\bibitem{zhang2004mwb}
J.~Zhang and M.~Fossorier, ``{A modified weighted bit-flipping decoding of
  low-density parity-check codes},'' \emph{IEEE Communications Letters},
  vol.~8, no.~3, pp. 165--167, 2004.

\bibitem{jiang2005imw}
M.~Jiang, C.~Zhao, Z.~Shi, and Y.~Chen, ``{An improvement on the modified
  weighted bit flipping decoding algorithm for LDPC codes},'' \emph{IEEE
  Communications Letters}, vol.~9, no.~9, pp. 814--816, 2005.

\bibitem{wu2007pwb}
X.~Wu, C.~Zhao, and X.~You, ``{Parallel Weighted Bit-Flipping Decoding},''
  \emph{IEEE Communications Letters}, vol.~11, no.~8, pp. 671--673, 2007.

\bibitem{ngatched2009ida}
T.~Ngatched, F.~Takawira, and M.~Bossert, ``{An improved decoding algorithm for
  finite-geometry LDPC codes},'' \emph{IEEE Transactions on Communications},
  vol.~57, no.~2, pp. 302--306, 2009.

\bibitem{li2008hid}
J.~Li and X.~Zhang, ``{Hybrid Iterative Decoding for Low-Density Parity-Check
  Codes Based on Finite Geometries},'' \emph{IEEE Communications Letters},
  vol.~12, no.~1, pp. 29--31, 2008.

\bibitem{li2009ipw}
G.~Li and G.~Feng, ``{Improved parallel weighted bit-flipping decoding
  algorithm for LDPC codes},'' \emph{IET Communications}, vol.~3, no.~1, pp.
  91--99, 2009.

\bibitem{wiberg1996cad}
N.~Wiberg, \emph{{Codes and decoding on general graphs}}.\hskip 1em plus 0.5em
  minus 0.4em\relax Department of Electrical Engineering, Link{\"o}ping
  University, 1996.

\bibitem{fossorier1999rci}
M.~Fossorier, M.~Mihaljevic, and H.~Imai, ``{Reduced complexity iterative
  decoding of low-density parity checkcodes based on belief propagation},''
  \emph{IEEE Transactions on Communications}, vol.~47, no.~5, pp. 673--680,
  1999.

\bibitem{chen2005rcd}
J.~Chen, A.~Dholakia, E.~Eleftheriou, M.~Fossorier, and X.~Hu,
  ``{Reduced-Complexity Decoding of LDPC Codes},'' \emph{IEEE Transactions on
  Communications}, vol.~53, no.~8, pp. 1288--1299, 2005.

\bibitem{ardakani2006gsd}
M.~Ardakani and F.~Kschischang, ``{Gear-shift decoding},'' \emph{IEEE
  Transactions on Communications}, vol.~54, no.~7, pp. 1235--1242, 2006.

\bibitem{storn1997des}
R.~Storn and K.~Price, ``{Differential evolution-a simple and efficient
  adaptive scheme for global optimization over continuous spaces},''
  \emph{Journal of Global Optimization}, vol.~11, no.~4, pp. 341--359, 1997.

\bibitem{xiaofu7tuw}
W.~Xiaofu, L.~Cong, J.~Ming, X.~Enyang, Z.~Chunming, and Y.~Xiaohu, ``{Towards
  understanding weighted bit-flipping decoding},'' in \emph{IEEE Int. Symp.
  Inform. Theory}, vol.~7, pp. 1561--1566.

\bibitem{levin2007lsl}
D.~Levin, E.~Sharon, and S.~Litsyn, ``{Lazy scheduling for LDPC decoding},''
  \emph{IEEE Communications Letters}, vol.~11, no.~1, pp. 70--72, 2007.

\bibitem{kfir2003pvs}
H.~Kfir and I.~Kanter, ``{Parallel versus sequential updating for belief
  propagation decoding},'' \emph{Physica A: Statistical Mechanics and its
  Applications}, vol. 330, no. 1-2, pp. 259--270, 2003.

\bibitem{sharon2007esm}
E.~Sharon, S.~Litsyn, and J.~Goldberger, ``{Efficient Serial Message-Passing
  Schedules for LDPC Decoding},'' \emph{IEEE Transactions on Information
  Theory}, vol.~53, no.~11, pp. 4076--4091, 2007.

\end{thebibliography}
\end{document}